# Precursors of Gate Oxide Degradation in Silicon Carbide MOSFETs

Ujjwal Karki[*] and Fang Zheng Peng[*]

**ABSTRACT**

Gate oxide degradation is more critical in Silicon-Carbide (SiC) MOSFETs than in Silicon (Si) MOSFETs. This is because of the smaller gate oxide thickness and the higher electric field that develops across the gate oxide in SiC MOSFETs. While multiple precursors have been identified for monitoring the gate oxide degradation in Si MOSFETs, very few precursors have been identified for SiC MOSFETs. The purpose of this paper is to demonstrate that gate oxide degradation precursors used in Si MOSFETs: a) threshold voltage, b) gate plateau voltage and c) gate plateau time, can also be used as precursors for SiC MOSFETS. Moreover, all three precursors are found to exhibit a simultaneous increasing trend (during the stress time) leading to an increase in on-state loss, switching loss and switching time of the SiC MOSFET. The existing studies of gate oxide degradation mechanisms in SiC MOSFETs, and their effects on threshold voltage and mobility were extended to correlate a variation of all three precursors using analytical expressions. The increasing trends of precursors were experimentally confirmed by inducing gate oxide degradation in commercial SiC MOSFET samples.

## I. INTRODUCTION

The gate oxide material in both Si and SiC MOSFETs is predominantly silicon dioxide ($SiO_2$). This gate oxide material degrades over time and the degradation process is more critical in SiC MOSFETs than in Si MOSFETs. This is because the electric field developed in the $SiO_2$ in SiC material (nearly 2.5 times the breakdown strength of SiC i.e., 2.5 x 3 MV/cm) is approximately ten times larger than the electric field developed in the SiO2 in Si material (nearly 3 times the breakdown strength of Si i.e., 3 x 0.25 MV/cm) [1]–[3]. Moreover, the gate oxide thickness is smaller in case of SiC MOSFETs [3], [4]. As a result, the gate oxide could easily reach its reliability limits; therefore, it is extremely important to monitor the effects of gate oxide degradation process in SiC MOSFETs.

Gate oxide degradation significantly alters the electrical parameters of power MOSFETs; to observe this, the electrical parameters are utilized as precursors of gate oxide degradation. Multiple precursors of gate oxide degradation have been identified for Si MOSFETs. The threshold voltage ($V_{TH}$) is the most common precursor for Si MOSFETs [5]–[7] . Very recent literature have also reported the gate plateau voltage ($V_{GP}$) [6] and the gate plateau time ($t_{GP}$) [7] as new online precursors of gate oxide degradation in Si MOSFETs. On the other hand, only the threshold voltage [8], [9] and the gate leakage current [10], [11] have been reported as precursors of gate oxide degradation in SiC power MOSFETs. Though the oxide degradation mechanism is different in Si and SiC MOSFETs, we demonstrate that three precursors used in Si MOSFETs: $V_{TH}$, $V_{GP}$ and $t_{GP}$, are also suitable for monitoring gate oxide degradation in SiC MOSFETs. It is interesting to note that the extraction of these precursors: $V_{TH}$, $V_{GP}$ and $t_{GP}$, can all be done from the same turn-on waveform of the power MOSFET as shown in Fig. 1.

The gate oxide degradation mechanism occurring within Si-SiO2 structures and SiC-SiO2 structures are different. In Si MOSFETs, the gate oxide degradation results in the formation of positive oxide-trapped charges within the gate oxide (which leads to a decrease of precursor magnitude) and negative interface-trapped charges at the oxide-silicon interface (which leads to an increase of precursor magnitude) [5], [7]. All three precursors have been found to exhibit a simultaneous dip-and-rebound variation pattern over time [7]. In contrast, the gate oxide degradation in SiC MOSFETs occurs due to the direct tunneling of electrons into the preexisting oxygen vacancy defects that exist near the SiC-$SiO_2$ interface (also called near-interface oxide traps) [8], [9], [12], [13]. The electrons tunneling into the near-interface oxide traps ($Q_{not}$) is dependent upon the bias-stress magnitude and the bias-stress time. In other words, the longer the bias-stress time (or the higher bias-stress), the deeper into the oxide the electrons can tunnel into. As a result, upon positive gate bias stress, the threshold voltage in SiC MOSFETs are found to increase with time [8], [9], [14]. It is important to note that both $V_{GP}$ and $t_{GP}$ bear a strong analytical relationship with $Q_{not}$ through the threshold voltage (to be discussed later). Therefore, we can expect both $V_{GP}$ and $t_{GP}$ to exhibit a simultaneous increasing trend like $V_{TH}$ (during the stress time) as shown in Fig. 2.

[*]The authors are with the Department of Electrical and Computer Engineering, Michigan State University, East Lansing, MI 48824 USA (e-mail: karkiujj@egr.msu.edu; fzpeng@egr.msu.edu).

The increasing trend of electrical parameters due to gate oxide degradation can be detrimental to device performance, reliability and efficiency. The degradation causes a simultaneous increase of $V_{TH}$, $V_{GP}$ and $t_{GP}$, which correspond to an increase in on-state loss, switching loss and switching time of the MOSFET, respectively. A holistic approach inclusive of the variation of all three precursors is, therefore, necessary to understand the detrimental effects of gate oxide degradation in SiC power MOSFETs.

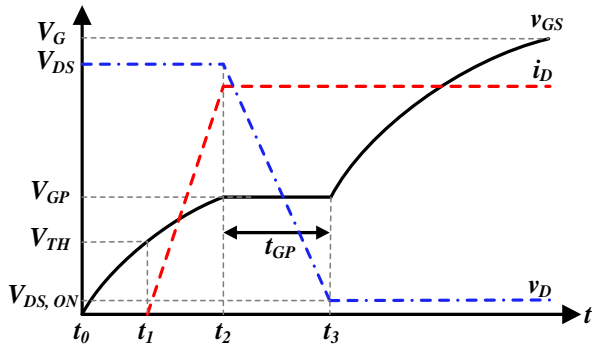
Fig. 1. Typical turn-on waveform of Power MOSFETs

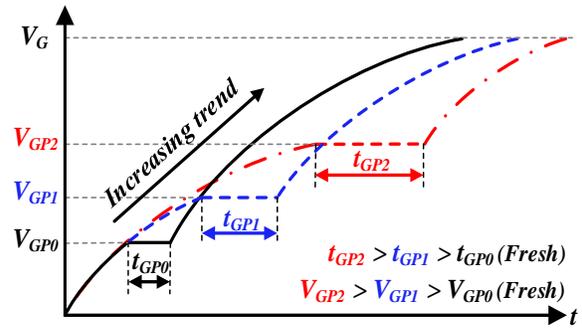
Fig. 2. Expected variation of precursors over stress-time

The organization of the paper is as follows. Section II provides background information about the effect of gate oxide degradation in SiC MOSFETs. Section III investigates the variation of all three precursors due to charge trapping in near-interface oxide traps. The subsequent sections provide experimental verification of increasing variation trends of precursors in a commercial 1200 V, 36 A SiC MOSFET.

## II. GATE OXIDE DEGRADATION MECHANISM IN SIC MOSFETS

The number of oxygen vacancy related defects in SiO2 is much larger in SiO2-SiC structure than in Si-SiO2 structure [15]–[17]. These oxygen vacancy related defects introduce defect energy bands ($E_T$) near the conduction band of SiC as sketched in Fig. 3a (dotted lines). Furthermore, these defects are electrically active defects in that they can participate in the charge trapping process, thereby affecting the electrical parameters. Because these defects are located near the interface of SiC and SiO2 structure, and are able to trap charge carriers, these defects are also called near-interfacial oxide traps ($Q_{not}$) [8], [12].

The electron trapping mechanism in $Q_{not}$ is explained next. When a positive bias is applied to turn on the MOSFET, the applied electric field causes the surface to invert. As a result, a large number of electrons accumulate near the conduction band edge of SiC. These electrons, which are in excited state, can then tunnel into the defect energy bands (which are lower in energy level compared to conduction band edge of SiC) and occupy the oxide traps as shown in Fig. 3. The number of electrons tunneling into $Q_{not}$ depends upon the duration and magnitude of positive stress; a longer stress duration causes electrons to penetrate deeper into the oxide [13], [18].

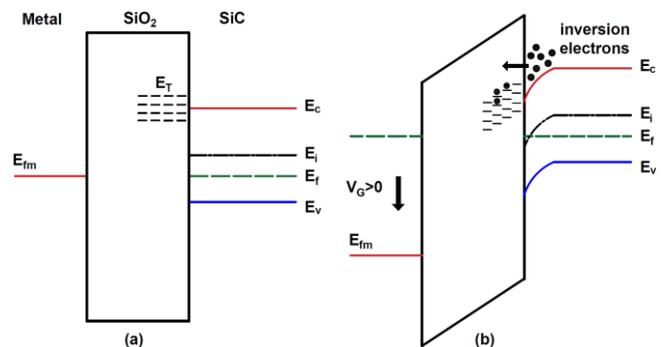
Fig. 3. a) Energy band diagram of a SiC/SiO2 structure with a defect band, and b) Electron tunneling into oxide traps under positive gate bias

The electrons that become trapped in these oxide traps creates a negatively charged $Q_{not}$ which opposes the effect of applied oxide field. Thus, a higher electric field is required across the oxide for channel inversion. This leads to an increase in threshold voltage, which is given by [12] :

$$V_{TH} = V_{TH0} + \frac{qN_{not}}{C_{ox}}, \tag{1}$$

where $V_{TH0}$ is the initial value of threshold voltage and $N_{not}$ is the number of $Q_{not}$s participating in the electron trapping process. The threshold voltage increases because of increasing number of active $N_{not}$s (traps that have captured electrons) during gate-bias stress.

## III. INVESTIGATION OF VARIATION OF PRECURSORS

In this section, the analytical expressions of three precursors are analyzed to show their respective variations due to increasing number of active near-interface oxide traps. The structure of both Si and SiC MOSFETs are similar. Thus, basic expression of electrical parameters ($V_{TH}$, $V_{GP}$ and $t_{GP}$) are the same in both.

### A. Variation of threshold voltage

The partial derivative of $V_{TH}$ in (1) with respect to (w.r.t) $N_{not}$ results in:

$$\frac{\partial V_{TH}}{\partial N_{not}} = \frac{q}{C_{ox}} > 0. \tag{2}$$

The partial derivative of $V_{TH}$ w.r.t $N_{not}$ is positive. This follows our previous discussion that $V_{TH}$ increases with increase of $N_{not}$. It is also important to mention that electron trapping in $Q_{not}$ reduces channel carrier mobility (μ) by coulombic scattering of the channel charge carriers [12]. In other words,

$$\frac{\partial \mu}{\partial N_{not}} < 0 \tag{3}$$

### B. Variation of gate plateau voltage

The gate voltage remains constant (or flat) during the time span ($t_3$-$t_2$) of the turn-on characteristics shown in Fig 1. This is referred to as the gate plateau voltage (or miller platform voltage), and is given by [19]:

$$V_{GP} = V_{TH} + \sqrt{\frac{I_D L_{CH}}{\mu C_{ox} Z}}, \tag{4}$$

where $I_D$ is the drain current, $L_{CH}$ is the channel length, $Z$ is the channel width and $C_{ox}$ is the specific gate oxide capacitance.

The partial derivative of $V_{GP}$ with respect to (w.r.t) $N_{not}$ results in:

$$\frac{\partial V_{GP}}{\partial N_{not}} = \frac{\partial V_{TH}}{\partial N_{not}} - \frac{1}{2\mu^{3/2}} \sqrt{\frac{I_D L_{CH}}{C_{ox} Z}} \frac{\partial \mu}{\partial N_{not}} > 0. \tag{5}$$

Since $V_{TH}$ increases w.r.t $N_{not}$ and μ decreases w.r.t $N_{not}$, the partial derivative of $V_{GP}$ w.r.t $N_{not}$ in (5) is positive. Thus, $V_{GP}$ increases with increase of active $N_{not}$. The increasing trend of $V_{GP}$ is illustrated in Fig. 2, where $V_{GP2} > V_{GP1} > V_{GP0}$ (Fresh MOSFET).

### C. Variation of gate plateau time

The gate plateau time ($t_{GP}$) is the time span where the gate voltage remains constant and equal to $V_{GP}$ (time span ($t_3$-$t_2$) in Fig. 1). The time interval ($t_3$-$t_2$) of a power MOSFET is given by [19]:

$$t_{GP} = t_3 - t_2 = \frac{R_G C_{GD,av}}{V_G - V_{GP}}\left[V_{DS} - v_{D(t3)}\right]. \qquad (6)$$

As shown in the turn-on waveform in Fig. 1, at the end of time $t_3$, the drain voltage $v_D(t_3)$ becomes nearly equal to the on-state voltage drop of the power MOSFET. Thus, $v_D(t_3)$ can be considered negligible. Therefore, the expression for $t_{GP}$ in (6) can be re-written as:

$$t_{GP} = t_3 - t_2 = R_G C_{GD,av} \frac{V_{DS}}{V_G - V_{GP}} . \qquad (7)$$

The partial derivative of $t_{GP}$ w.r.t. $N_{not}$ results in:

$$\frac{\partial t_{GP}}{\partial N_{not}} = R_G C_{GD,av} \frac{V_{DS}}{(V_G - V_{GP})^2} \frac{\partial V_{GP}}{\partial N_{not}} > 0 . \qquad (8)$$

Since the partial derivative of $V_{GP}$ w.r.t $N_{not}$ is positive in (5), the partial derivative of $t_{GP}$ w.r.t $N_{not}$ in (8) is also positive. This indicates that $t_{GP}$ increases with increase of $N_{not}$. The increasing trend of $t_{GP}$ is illustrated in Fig. 2, where $t_{GP2} > t_{GP1} > t_{GP0}$ (Fresh MOSFET).

## IV. EXPERIMENT DETAILS

Experiments were performed on 1200 V, 36 A commercial SiC MOSFETs to validate the analysis made in previous section. The experimental setup mainly consists of a High Electric Field (HEF) stress platform (Fig. 4a) to induce accelerated gate oxide degradation in SiC MOSFETs and a switching-test platform (Fig. 4b) to obtain the turn-on waveform of SiC MOSFETs.

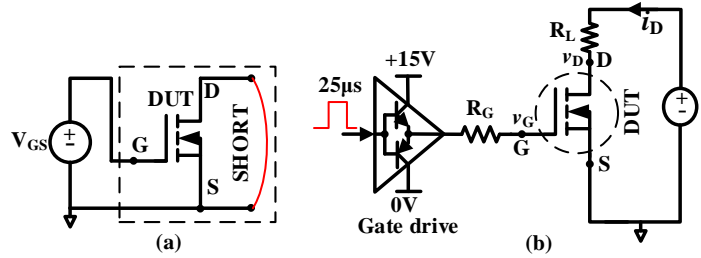

Fig. 4. Circuit schematic for: a) HEF-stress platform, and b) Switching-test platform

The accelerated aging of gate oxide was induced by applying positive bias to the gate electrode at room temperature, with the drain and source terminals grounded. A safe stress level of +45 V was chosen such that it was sufficient to initiate observable gate oxide degradation, but low enough not to cause gate oxide breakdown during the experiment.

To obtain the turn-on waveform, each MOSFET was driven by a gate driver with a gate bias voltage of 0 V/+15 V (see Fig. 4b). Please note that a gate voltage of 0 V (instead of a negative gate voltage) was used to prevent the oxide traps from discharging the trapped electrons during the off-state. This enables us to observe the full extent of gate oxide degradation. Since, it is easier to observe switching characteristics with a larger gate resistance, an external 100 Ω resistance was inserted between the gate driver and the gate terminal of the MOSFET. To minimize self-heating of the MOSFET during the test, a single 25 μs pulse was provided to the gate driver. A DC voltage of 30 V and a load resistance ($R_L$) of 30 Ω were used to ensure a drain current of 1 A.

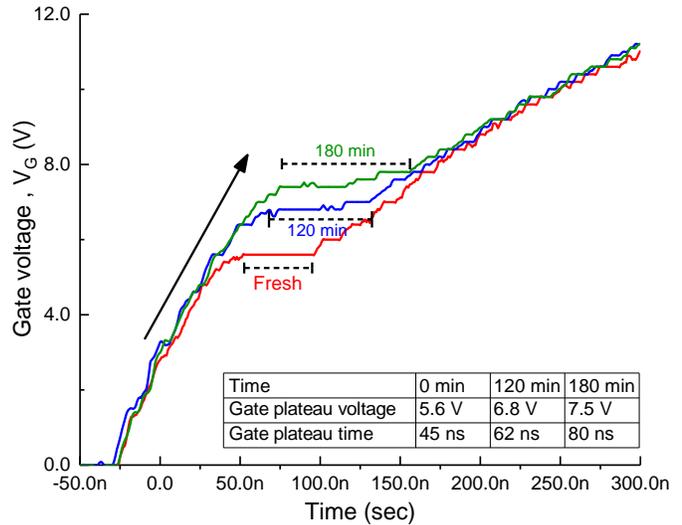

| Time | 0 min | 120 min | 180 min |
|---|---|---|---|
| Gate plateau voltage | 5.6 V | 6.8 V | 7.5 V |
| Gate plateau time | 45 ns | 62 ns | 80 ns |

Fig. 5. Variation of $V_{GP}$ and $t_{GP}$ shown in the gate voltage waveform

## V. RESULTS

Fig. 5 shows gate voltage ($V_G$) waveforms for a 1200 V, 36 A commercial SiC MOSFET sample. It is seen that both $V_{GP}$ and $t_{GP}$ increase over the duration of stress. The solid black arrow indicates an increasing trend of $V_{GP}$ with degradation. Tabulated $V_{GP}$ values show that $V_{GP}$ @ 180 min > $V_{GP}$ @ 120 min > $V_{GP}$ @ 0 min (Fresh MOSFET). Similarly, the dotted black lines show $t_{GP}$ during different stress times. Tabulated $t_{GP}$ values show that $t_{GP}$ @ 180 min > $t_{GP}$ @ 120 min > $t_{GP}$ @ 0 min (Fresh MOSFET). The results confirmed that both $V_{GP}$ and $t_{GP}$ reflected the gate oxide degradation status with simultaneous positive shifts from their initial values.

## VI. CONCLUSION

In this paper, the effect of gate oxide degradation on three electrical parameters of SiC MOSFETs were examined. It was demonstrated that the electrical parameters of SiC MOSFETs i.e., $V_{TH}$, $V_{GP}$ and $t_{GP}$ exhibit increasing trends over the course of gate oxide degradation and can be considered as effective precursors of gate oxide degradation in SiC MOSFETs. The increasing trend, which results from the electron trapping in near-interface oxide traps, was confirmed by inducing gate-oxide degradation in commercial SiC MOSFETs.